\documentclass[prl,amssymb,twocolumn,aps,floatfix,showpacs]{revtex4-1}
\usepackage{graphicx}

\usepackage{epstopdf}
\begin{document}

\newif\ifShow
\newif\ifMethodone

\Methodonefalse
\Showfalse

\title{Spin and spatial dynamics in electron-impact scattering off S-wave
He using R-matrix with Time-Dependence theory}
\author{Jack Wragg and Hugo W. van der  Hart}

\affiliation{Centre for Theoretical Atomic, Molecular and Optical Physics, School of Mathematics and Physics,
Queen's University Belfast, Belfast, BT7 1NN, United Kingdom}

\begin{abstract}
R-matrix with time-dependence theory is applied to electron-impact ionisation processes for He in the
S-wave model. Cross sections for electron-impact excitation, ionisation and
ionisation with excitation for impact energies between 25 and 225 eV
are in excellent agreement with benchmark cross sections. Ultra-fast dynamics
induced by a scattering event is observed
through time-dependent signatures associated with autoionisation from doubly excited
states. 
Further insight into dynamics can be obtained through examination
of the spin components of the time-dependent wavefunction.\end{abstract}

\maketitle

\section{Introduction}

Scientific progress greatly benefits from the development of theoretical and computational methods that complement new experimental techniques. Recent
developments in the study of electron dynamics on the sub-femtosecond timescale
\cite{Sansone2010,Schultze2013,Schiffrin2013,Popmintchev2012} have enhanced the
need for the development of  computational models able to obtain a time-dependent description
of ultrafast multi-electron processes. In the present manuscript, we demonstrate a new
time-dependent ab-initio
computational method for the study of electron spatial and spin dynamics through its application to
electron-He impact processes in the S-wave model (known as the Temkin-Poet model when applied to
electron-hydrogen scattering \cite{temkin1961,Poet1978}). We choose this particular model as it provides a simple atomic process which contains both spin and
spatial dynamics, and for which benchmark data for comparison is readily available
\cite{Bartlett2010,Bartlett2010II}. 

Electron-impact processes for He in the S-wave model were investigated through application of the
time-dependent Close-Coupling (TDCC) approach \cite{Pindzola1999}. Since then, a range of other
advanced approaches have been applied to investigate this problem, including the convergent
close-coupling approach \cite{Fursa1995, plottke2002,plottke2004} and the exterior-complex-scaling
approach (ECS) \cite{Horner2005,Bartlett2010,Bartlett2010II}. We note that this description of a
three-electron system bears great similarity to the time-dependent calculation of Li processes in
\cite{Ruiz2005}, since the restriction in angular momentum corresponds to a 1D description for each
electron.


In this report, we build upon the R-Matrix with Time-dependence (RMT) theory
\cite{Lysaght2012,Nikolopoulos2008}. This approach
combines the R-matrix division of configuration space with time propagation to model
attosecond processes in many-electron
atoms. RMT has recently provided valuable insights into high-harmonic generation \cite{Hassouneh2014},
and experimental attosecond transient absorption spectroscopy
data \cite{Ding2016}. The RMT approach has
been extended to model dynamics in atomic systems where two electrons
are ejected from the core, demonstrated with an application to
double photoionisation from a Helium atom \cite{Wragg2015}. 


We use RMT theory to consider ultra-fast dynamics that occur
within electron-impact excitation,
ionisation, and in particular ionisation-excitation processes. Whereas previous application of RMT theory for
two electron ejection considered systems with a single double-ejection threshold, the present study
investigated the numerical accuracy of the approach for systems with multiple
thresholds. The present process provides an opportunity to robustly assess the numerical techniques in the RMT approach through quantitative comparison with present data for electron-impact of He in the S-wave model  
\cite{plottke2002,plottke2004,Horner2005,Bartlett2010,Bartlett2010II}. 

The treatment of this particular problem provides a stepping stone towards the development of an RMT approach for the full treatment of double ionisation in general atomic systems. The backbone of the treatment would be formed by states consisting of a double ionisation threshold two or more free electrons. The electron-He scattering process can be regarded as the simplest of such systems. The RMT approach for this scattering process thus offers a clear development path towards a general atomic code for the treatment of double ionisation. In addition to this, the time-dependent nature of the RMT treatment can provide dynamical insight into the scattering processes. For example, whereas the excitation of
autoionizing states in this scattering process has been considered previously \cite{Horner2005},
a time-dependent treatment can reveal clear signatures of dynamics within these states. Furthermore, a time-dependent method, such as RMT also allows the spin coupling between the electrons to be traced
during the scattering process.

Throughout this paper, we use atomic units unless otherwise stated.

\section{Theory}
In RMT, the three-electron S-wave Hamiltonian
\begin{equation}
\hat{H}=\sum_{n=1}^{3}\left(-\frac{1}{2}\frac{d^{2}}{dr_{n}^{2}}-\frac{2}{r_{n}}\right)+
\frac{1}{r^{>}_{12}}+\frac{1}{r^{>}_{13}}+\frac{1}{r^{>}_{23}} ,
  \label{fullham}
\end{equation}
is used within the Schr\"odinger equation
\begin{equation}
  i\frac{d}{dt}\Psi(\vec{R},\chi,t)=\hat{H}\Psi(\vec{R},\chi,t)
\end{equation}
where $\vec{R}$ is the position vector $(r_{1},r_{2},r_{3})$, and $r_{n}$ is the radial coordinate of electron $n$. $r^{>}_{n'n''}$
is the greater of $r_{n'}$ and $r_{n''}$. $\Psi(\vec{R},\chi,t)$ is the time-dependent wavefunction where $\chi$ indicates the spin coupling of the electrons.

Three regions of configuration space are defined within RMT for two-electron ejection \cite{Wragg2015}: region (I) with $r_{1},r_{2},r_{3}<b$, where $b$ is the size of the
so-called inner region, region (II) where $r_{1},r_{2}<b$ and $r_{3}>b$, and region (III)
where $r_{1}<b$ and $r_{2},r_{3}>b$. The RMT wavefunction in each region is described
in terms of a time-dependent coefficient and a time-independent basis as
\begin{eqnarray}
&\text{(I) } & \Psi(\vec{R},\chi,t)= \sum_{j}C^{\mathrm{(I)}}_{j}(t)\psi^{\mathrm{(I)}}_{j}(r_{1},r_{2},r_{3},\chi) \\
&\text{(II) } & \Psi(\vec{R},\chi,t)= \sum_{k}C^{\mathrm{(II)}}_{k}(r_{3},t)\psi^{\mathrm{(II)}}_{k}(r_{1},r_{2},\chi) \\
&\text{(III) } & \Psi(\vec{R},\chi,t)= \sum_{m}C^{\mathrm{(III)}}_{m}(r_{2},r_{3},t)\psi^{\mathrm{(III)}}_{m}(r_{1},\chi) 
\end{eqnarray}
where the coefficients $C^{\mathrm{(II)}}_{k}(r_{3},t)$, and $C^{\mathrm{(III)}}_{m}(r_{2},r_{3},t)$ are
defined at FD grid points across $r_{3} >b$, and $r_{2},r_{3}>b$ respectively. $k$ and $m$
correspond to single- and two-electron channels in regions (II) and (III) respectively, $j$ indicates
a region (I) eigenstate. Three-electron escape corresponding to $r_{1},r_{2},r_{3}>b$ is not
considered.
Configuration space not covered in regions (I), (II) and (III) is included via antisymmetrisation
of the wavefunction. 

The basis functions $\psi^{\mathrm{(N)}}$ are
expanded in terms of a further basis of functions, $\zeta^{\mathrm{(N)}}_{k}$ (for (N) = (I), (II), and (III)), and appropriate
spin functions. These $\zeta^{\mathrm{(N)}}_{k}$ functions are in turn constructed from antisymmetrised products
of hydrogenic eigenfunctions $\zeta^{\text{+}}_{n}(r_{i})$, corresponding to the
$n^{\rm th}$ eigenvalue of the operator
\begin{equation}
\hat{H}^{+}_{i} = -\frac{1}{2}\frac{\mathrm{d}^{2}}{\mathrm{d}r_{i}^{2}}-\frac{2}{r_{i}}+L_b,
\label{Ham+}
\end{equation} 
where $L_b$ is the Bloch operator \cite{burke2011}, written as
\begin{equation}
L_b=\frac{1}{2}\delta(r_i-b)\frac{d}{dr_i}.
\end{equation}
To minimise the number of basis functions, at least one of the electrons within the inner region $(r_i<b)$ is
restricted to the lowest three eigenstates. 
This ``core" electron is thus restricted to
the 1s, 2s and 3s orbitals. We obtain eigenfunctions for the inner-region $(r_i<b)$ aspect
 of the wavefunction in each of the three regions through diagonalisation
of the following Hamiltonians:
\begin{eqnarray}
& {\mathrm{(I)}} & \hspace{5pt} \hat{H}^{+}_{1}+ \hat{H}^{+}_{2} + \hat{H}^{+}_{3} +
\frac{1}{r^{>}_{12}} + \frac{1}{r^{>}_{23}} +\frac{1}{r^{>}_{13}} \nonumber \\
& {\mathrm{(II)}} &\hspace{5pt}   \hat{H}^{+}_{1}+ \hat{H}^{+}_{2} + \frac{1}{r^{>}_{12}} \label{operators} \\
& {\mathrm{(III)}} &\hspace{5pt}    \hat{H}^{+}_{1}
\nonumber
\end{eqnarray}
where 
$\hat{H}^{+}_{i}$ is the hydrogenic Hamiltonian given in equation \ref{Ham+}. 

As with previous RMT implementations, the wavefunction is propagated in time from an initial state at
$t=t_{0}$. 
This initial state contains two electrons in the He ground state, and an incoming $s$ electron, described by a Gaussian wavepacket of $\text{root-mean-square width}=10\ a_{0}$ centred on $r_{3}=75\ a_{0}$.

 In this study, we use a 6$^{\rm th}$ order Taylor series propagator.
The kinetic-energy operations on the coefficients defined across FD grids $\big($$-\frac{1}{2}\frac{d^{2}}{dr_{3}^{2}}$ in region (II) and $-\frac{1}{2}\frac{d^{2}}{dr_{2}^{2}},-\frac{1}{2}\frac{d^{2}}{dr_{3}^{2}}$ in region (III)$\big)$ are evaluated using FD operators.
Near the inner boundary of regions (II) and (III), the FD grids contain insufficient grid points
to complete the centre difference FD operation. The missing data points are hence obtained from the
wavefunction in region (I) or region (II), respectively.
Additionally, propagation using the physical three-electron Hamiltonian in equation (\ref{fullham})
requires cancellation of the Bloch
operator terms contained within $\hat{H}_{3}^+$ (region I) and $\hat{H}_{2}^+$ (region II), as defined
in equation (\ref{operators}). This is achieved through the evaluation of an FD first
derivative operation on the wavefunction at the region (I)/region (II) boundary, and the region (II)/region (III) boundary,
as implemented in \cite{Wragg2015}.

\begin{figure}[t!]
	\begin{center}
		\includegraphics[width=0.5\textwidth,angle=0]{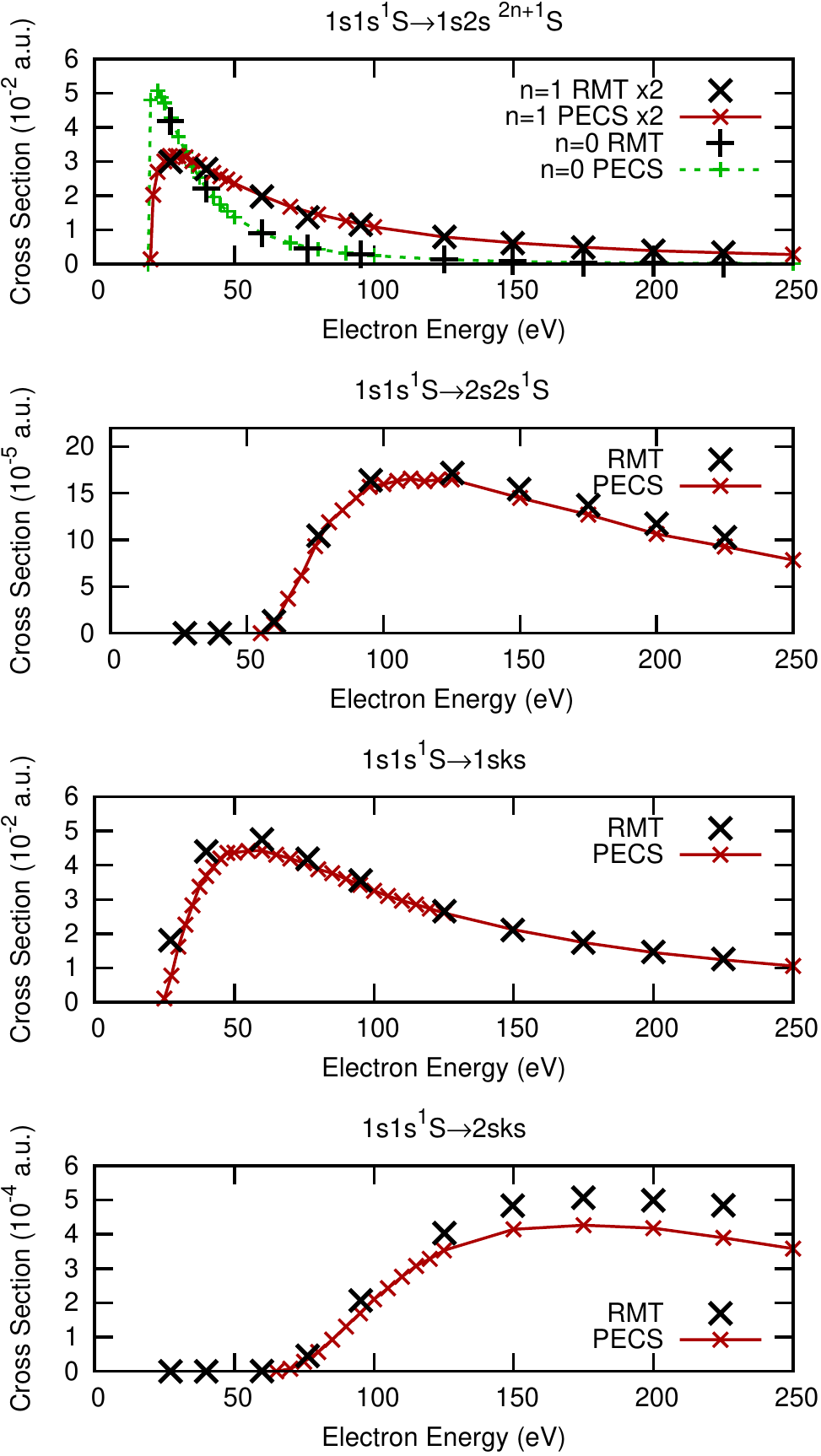}
		\caption{Electron-impact cross sections for He in the S-wave model for impact energies between
		25 and 225 eV as obtained in the RMT approach. Cross sections for
		electron-impact excitation to 1s2s, and 2s2s, electron-impact ionisation, leaving He$^+$ in 1s, and
		electron-impact ionisation with excitation of He$^+$ to 2s. All cross sections are compared with
		benchmark data (calculated using the PECS method) from \cite{Bartlett2010,Bartlett2010II}.}
		\label{CSPlot1}
	\end{center}
\end{figure}


The initial wavefunction is propagated in time until the scattered electron has moved well away from
the residual atom or ion. Electron-impact excitation yields are then obtained
from the total population in the relevant region (II) channel.
Electron-impact ionisation yields are obtained from the total population in region (III) associated with a
particular residual He$^+$ state (1s, 2s or 3s).
These yields are then transformed into electron-impact scattering cross sections.

\section{Results}

Figure \ref{CSPlot1} shows impact excitation and ionisation cross sections for He in the
S-wave model over the electron-impact energy range between 25 and 225 eV. For all processes shown, we observe good agreement with the benchmark data. The
largest difference (25\%) is seen at large impact energies for electron-impact ionisation with excitation of the
residual ion to the 2s state, where the restriction of the core electron to 1s, 2s or 3s
could have a more significant effect on the modelling. We also note a more pronounced difference near the
threshold for electron-impact ionisation. In this energy range, the main difficulty lies in
distinguishing two-electron ejection from the excitation of high-lying excited states
(a similar challenge was encountered in \cite{Wragg2015}). Overall, the cross sections for electron-impact scattering
show excellent agreement with those obtained in \cite{Bartlett2010}, and demonstrate the accuracy
of the present approach.

\begin{figure}[t!]
	\begin{center}
		\includegraphics[width=0.5\textwidth,angle=0]{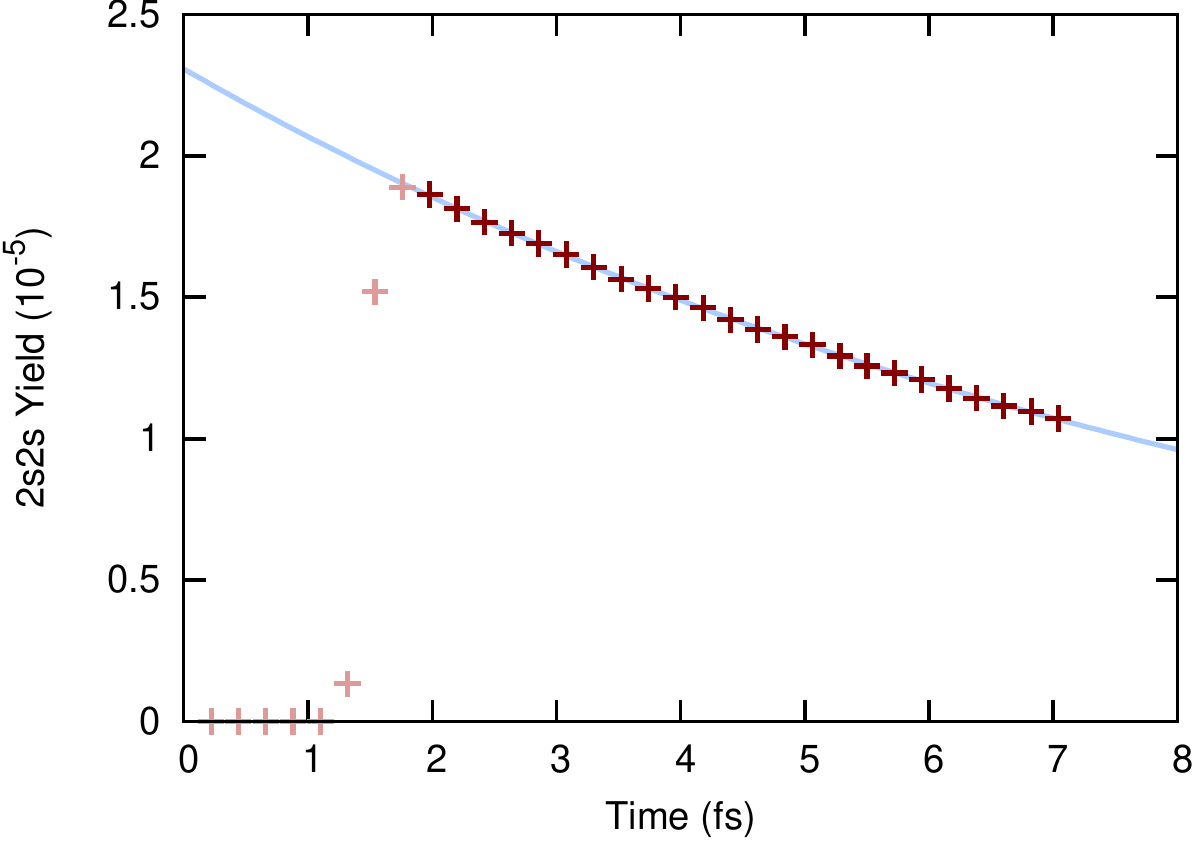}
		\caption{Yield of 2s2s state in region (II). Light red data points indicate a yield obtained before the excited wavepacket has entered region (II), and dark red data points indicate a yield obtained after the wavepacket has entered region (II). The blue line indicates an exponential decay fit of the dark red points.}
		\label{2s2s}
	\end{center}
\end{figure}

Following excitation, the population of the 2s2s state decreases over time as the state
autoionises. In figure \ref{2s2s}, the population of the channel associated with the 2s2s state in region (II) is shown as the calculation propagates in time. We note that the sharp increase in yield (light red points) corresponds not to the excitation of the 2s2s state, but rather to the flow of the scattered electron in the 2s2s channel from region (I) into region (II). The exponential decay of the 2s2s state is then seen in figure \ref{2s2s} (dark red points). The light blue line is fit of the exponential decay function $A\exp(-\gamma t)$ to the dark red points. From this fit, a 2s2s decay rate of $\gamma=1.10 \times 10^{14}$ s$^{-1}$ is obtained, which agrees to within 10\% with the decay rate given in \cite{Bartlett2010}. The earliest dark red point is taken as the yield for the 2s2s state, from which the value for the 2s2s cross sections shown in figure \ref{CSPlot1} are calculated. The moment of collision in the calculation shown in figure \ref{2s2s} is estimated to happen approximately $1.1$ fs after the beginning of the calculation, with the first reliable 2s2s yield obtained approximately $0.5$ fs later. We estimate that this lack of access to an immediate 2s2s yield introduces an uncertainty of approximately 10\% to the RMT 2s2s cross section. 

\begin{figure}[t!]
	\begin{center}
		\includegraphics[width=0.5\textwidth,angle=0]{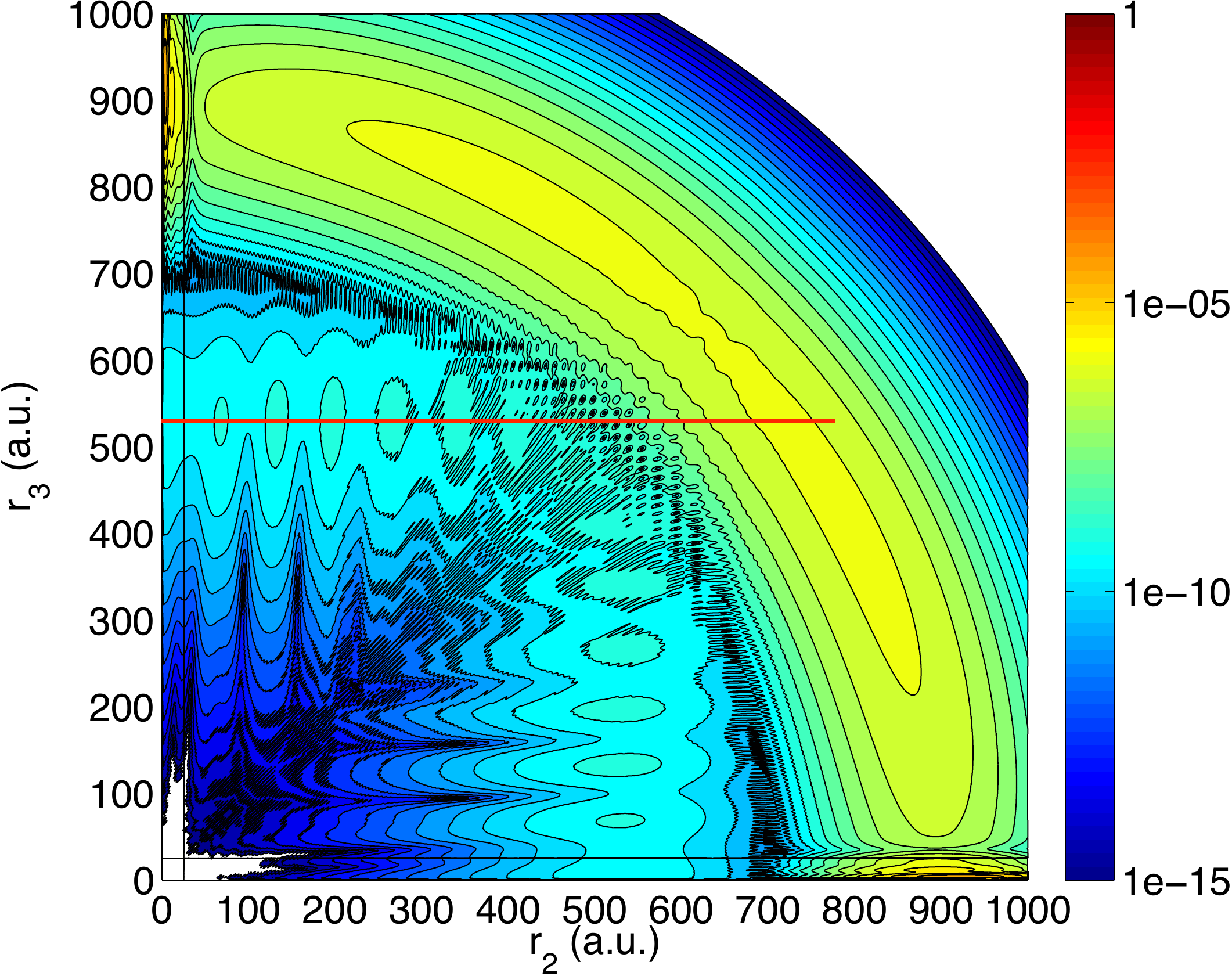}
		\caption{Probability density of finding electrons 2 and 3 at position $r_{2}$ and $r_{3}$ when
		the residual He$^+$ ion in the S-wave model is left in the 1s state. The density
		shown is obtained after 11.96 fs for an incoming electron wavepacket of 76 eV, initially centred at
		75 $a_0$. The red line at $r_{3}=535 a_{0}$ indicates data shown in figure \ref{Model}, 
		associated with autoionisation of the 2sns states.}
		\label{AutoIonisationPlot}
	\end{center}
\end{figure}

Previous studies have shown the theoretical time-dependent description of autoionisation to be an interesting challenge \cite{Schultz1994,Mitnik2004,Hu2005}. We can identify such autoionisation dynamics within the RMT model of the scattering process. We show in figure \ref{AutoIonisationPlot} the probability density
associated with a residual He$^+$ ion in the 1s state at 11.96 fs after the beginning of the model. The direct electron-impact-ionisation wavepacket
can be seen as an arc from $\sqrt{r_{2}^{2}+r_{3}^{2}}\approx 750$ $a_0$ to 
$\sqrt{r_{2}^{2}+r_{3}^{2}}\approx 1050$ $a_0$. In addition to this arc, a series of six peaks is seen along
along $r_{3}\approx 535 a_{0}$ and $r_{2}\approx 535 a_{0}$. These peaks signify
dynamics associated with doubly excited 2sns states and their autoionisation. The different nature of the two processes is reflected in the strong inteference where this series and arc overlap. We consider that the distance along coordinate $r_3$ at time $t$ corresponds to the momentum of a scattered electron after excitation of a doubly-excited 2sns state, i.e. $r_{3}=(t-t_{c})/k_{i_{n}}$. Here $t_{c}$ is value of $t$ at the moment of collision, and $k_{i_{n}}$ is the momentum of the impact electron after exciting the atom to the 2sns autoionising state.
Some time after the moment of collision, the 2sns state autoionizes, leading to emission of an
electron associated with the $r_2$ radial coordinate. Since this autoionised electron has a well-defined momentum,
$r_2$ and $t$ can be mapped to the moment of autoionisation $\tau_{n}(r_2,t)$ from state 2s$n$s as
\begin{equation}
\tau_{n}(r_2,t)=t - t_{c} - r_2/\sqrt{2 E_n}.
\end{equation}
such that the autoionised electron has travelled from $r=0$ at time of emission $\tau_{n}(r_2,t)$ to $r=r_{2}$ at time $t$. Interference between autoionisation contributions from 2s2s and higher 2sns states then gives a time-dependent autoionisation rate for the
doubly excited He atom, which is reflected in the series of peaks in figure \ref{AutoIonisationPlot}.

\begin{figure}[t!]
        \includegraphics[width=0.5\textwidth]{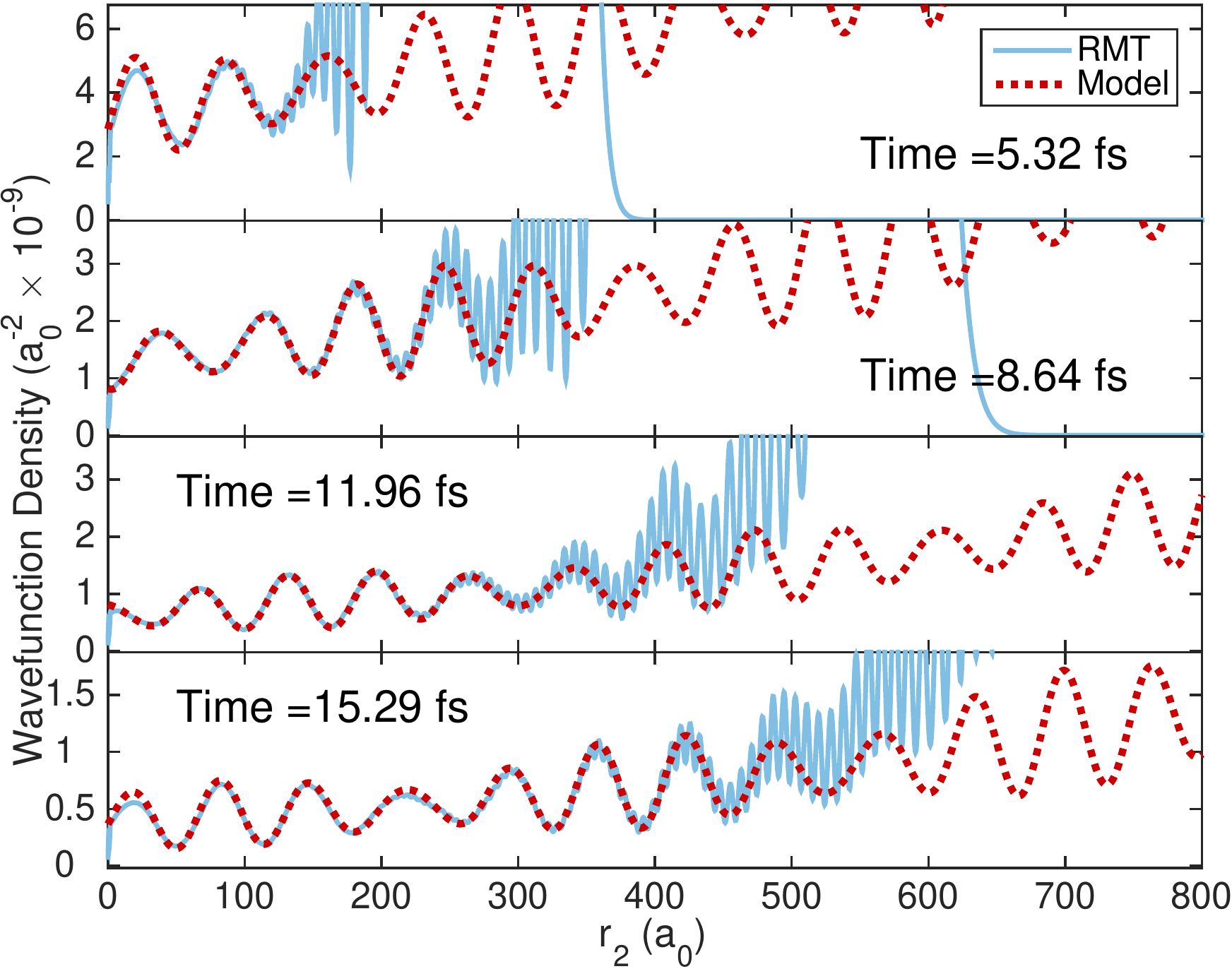}
        \caption{Wavefunction density along the lines $r_3=704 a_0, 535 a_0, 375 a_0, 217 a_0$ at times 15.29 fs, 11.96 fs, 8.64 fs, and 5.32 fs after the start of the model respectively. The 11.96 fs data corresponds to the data shown in figure \ref{AutoIonisationPlot}.
	 The RMT density (blue) is compared to model data (red), obtained as described in the text. The variations in
	 the density correspond to time variation in the autoionisation rate of a superposition of 2sns states.}
        \label{Model}
\end{figure}

Figure \ref{Model} shows the probability density along the autoionisation wavepacket (as described by the red line in figure \ref{AutoIonisationPlot}) at four moments during the calculation along with a model
of the autoionisation arising from the 2sns states. The
wavepacket along $r_2$ is modelled by considering the
autoionisation of the 2s2s, 2s3s and 2s4s states:
\begin{equation}
P(r_2,t) = C \left|\sum_{n=2}^4 \sqrt{\sigma_n \gamma_n e^{-  \gamma_n \tau_{n}(r_2,t)}}
  \exp {\left( i(k_{n}r_{2}-E_{n}t)\right)} \right|^2
\end{equation}
In these equations, $E_n$ is the energy of the 2sns state, $\gamma_n$ its autoionisation rate, and
$\sigma_n$ its cross section.
These quantities are obtained from \cite{Bartlett2010}. It is possible to obtain initial estimates for the normalisation constant $C$ from the shape of the R-Matrix wavepacket. However, for the sake of avoiding unnecessary complication, we obtain $C$ by a fit to each RMT dataset. Figure \ref{Model} shows
close agreement between the model and the RMT density. The rapid oscillations in figure \ref{Model} with a wavelength of $\approx 55a_{0}$ are related to interference between autoionisation from  2s2s and from the superposition of 2s3s and 2s4s. The modulation of these oscillations with a wavelength of $\approx 270 a_{0}$ are associated with the interference between autoionisation from 2s3s and 2s4s. Hence the sequence of peaks follows the time-varying
autoionisation of the doubly excited residual He atom.


We now turn to a demonstration of the capability of the RMT approach to describe spin dynamics as
well as spatial
dynamics. We note that RMT does not currently directly solve the relativistic Pauli or Dirac
equations as the regimes of interest here are non-relativistic. Rather, the changes in spin
coupling within the three electron system are inferred from the anti-symmetry that is imposed on
the wavefunction.

To illustrate how this kind of spin dynamics can manifest itself, we consider a simple thought
experiment of sequential double photoionisation of a spin-polarised three electron system (such
as atomic Li). An incoming high-energy photon can eject a 1s electron from the spin-polarised Li
1s$^{2}$2s ground state. The resulting 1s2s state will be in a superposition of 1s2s $^{1}$S and
1s2s $^{3}$S. The $m_S=\pm1$ components of this state can only be formed by the 1s2s $^3$S state.
However, the $m_{S}=0$ component of this state created by photoionization consists of a coherent
superposition of the $^1$S and $^3$S states. This superposition will now change over time between
$\left|1s\uparrow 2s\downarrow \right>$ and $\left|1s \downarrow 2s \uparrow \right>$, due to the
energy gap between the $^{1}$S and $^{3}$S states. Subsequent photoionisation of the 2s electron
by a short time-delayed pulse will then result in an observable time variation in the spin
polarisation of the ejected electron, signifying spin dynamics.

\begin{figure}[t!]
	\begin{center}
		\includegraphics[width=0.45\textwidth,angle=0]{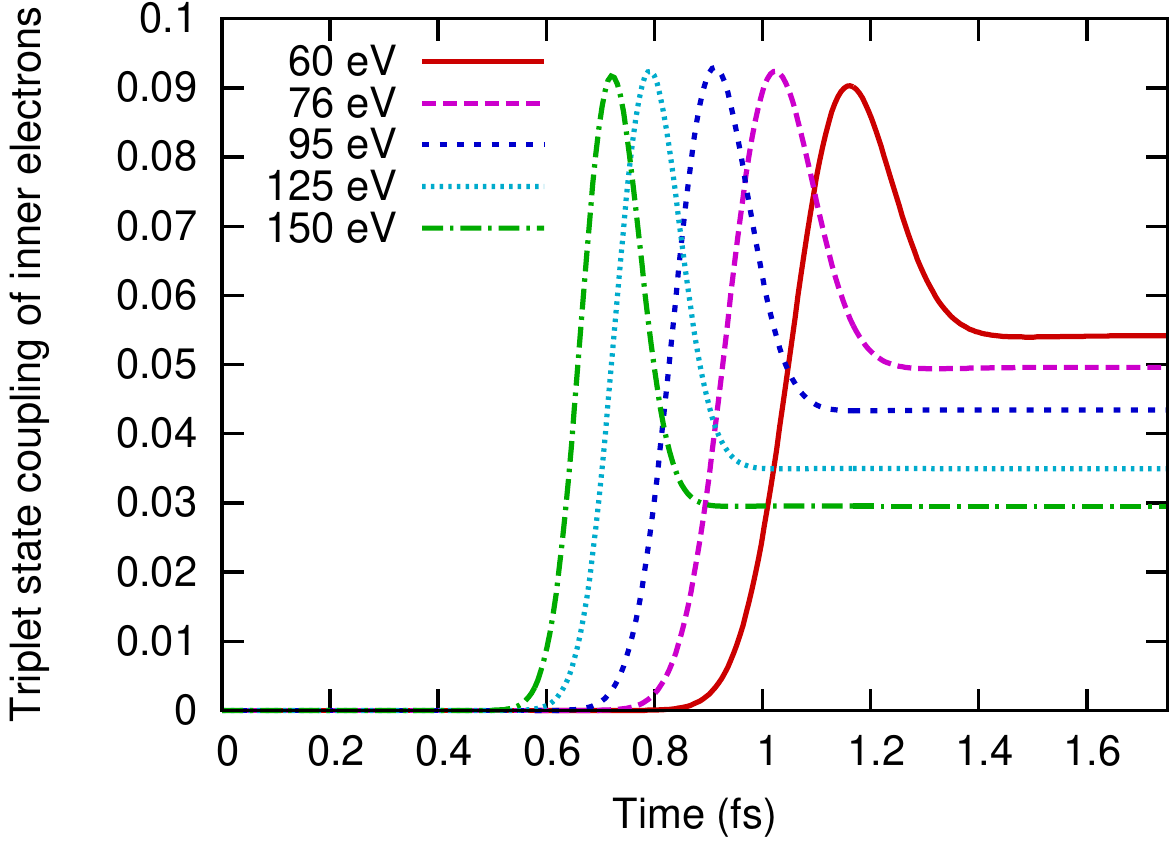}
		\caption{The fraction of the wavefunction in which the two innermost electrons are coupled
		to a triplet state as a function of time for different impact energies in electron
		scattering off He in the S-wave model. }
		\label{SpinDynamics}
	\end{center}
\end{figure}

The RMT approach offers the
capability to investigate such spin dynamics effects in an ab-initio manner.
This is demonstrated in 
figure \ref{SpinDynamics}, which shows the fraction of the three-electron wavefunction in which the innermost
two electrons are coupled to triplet spin symmetry as a function of time
for different electron-impact energies. Before the collision occurs, the innermost electrons are coupled to a singlet as the He atom is in the initial 1s$^2$ ground state.
During the collision, the incoming electron partially penetrates the ground state atom, becoming one of the innermost electrons. The coupling between impact electron and the other inner electron is partially described by a triplet coupling, causing the triplet spin fraction of the inner electrons to increase. After the collision, there is a notable probability for the impact electron to leave the atom, causing the original atomic electrons to return to being the inner electrons. This explains the later increase in singlet coupling.

Figure \ref{SpinDynamics} suggests that, for our particular choice of initial wavepacket, the main spin
dynamics in this scattering process occurs on a timescale that is dependent on the impact energy.
We note that access to the full time-dependent wavefunction enables the use of different recoupling
schemes, so it is possible to investigate the full range of dynamics in spin coupling between
electrons. This may be of particular interest when more complex atoms with different residual-ion states,
e.g. Ne$^{2+}$, are investigated.

\section{Conclusion}

In summary, the RMT approach has been successfully applied to study dynamics on the attosecond timescale for three-electron
systems from first principles. We have demonstrated that the RMT approach can reliably describe impact ionisation
processes involving double continua associated with different ionisation thresholds. This includes processes where the incoming electron electron excites a superposition of doubly excited states, which leads to ultra-fast dynamics
in the subsequent autoionisation. The autoionisation rates in region (II) are in excellent agreement with benchmark calculations. With RMT, it is possible to extract both spin and spatial dynamics from a single calculation. The RMT codes hence provide a foundation for the investigation of intense-field multiple ionisation processes
in three electron systems, as a stepping stone for our long-term aim to study such processes in general atoms.

This research was sponsored by the Engineering and Physical Sciences Research Council (UK) under grant
ref. no. G/055416/1. This work also used the ARCHER UK National Supercomputing service (http://www.archer.ac.uk).
Data from figure 1 to figure 5 can be accessed via http://pure.qub.ac.uk/portal/en/datasets/search.html
	\bibliography{Three}
\end{document}